\documentstyle[12pt]{article}
\input{psfig.sty}
\newlength{\dinwidth}
\newlength{\dinmargin}
\newlength{\extraspace}
\newlength{\extraspaces}
\newcommand{\be}{\begin{equation}
\addtolength{\abovedisplayskip}{\extraspaces}
\addtolength{\belowdisplayskip}{\extraspaces}
\addtolength{\abovedisplayshortskip}{\extraspace}
\addtolength{\belowdisplayshortskip}{\extraspace}}
\newcommand{\ee}{\end{equation}}
\newcommand{\bdm}{\begin{displaymath}
\addtolength{\abovedisplayskip}{\extraspaces}
\addtolength{\belowdisplayskip}{\extraspaces}
\addtolength{\abovedisplayshortskip}{\extraspace}
\addtolength{\belowdisplayshortskip}{\extraspace}}
\newcommand{\edm}{\end{displaymath}}
\renewcommand{\thefootnote}{\fnsymbol{footnote}}
\def\simlt{\mathrel{\lower2.5pt\vbox{\lineskip=0pt\baselineskip=0pt
           \hbox{$<$}\hbox{$\sim$}}}}
\def\simgt{\mathrel{\lower2.5pt\vbox{\lineskip=0pt\baselineskip=0pt
           \hbox{$>$}\hbox{$\sim$}}}}
\newcommand{\ls}[1]
   {\dimen0=\fontdimen6\the\font
    \lineskip=#1\dimen0
    \advance\lineskip.5\fontdimen5\the\font
    \advance\lineskip-\dimen0
    \lineskiplimit=.9\lineskip
    \baselineskip=\lineskip
    \advance\baselineskip\dimen0
    \normallineskip\lineskip
    \normallineskiplimit\lineskiplimit
    \normalbaselineskip\baselineskip
    \ignorespaces}

\catcode`@=11
\newcount\@tempcntc
\def\@citex[#1]#2{\if@filesw\immediate\write\@auxout{\string\citation{#2}}\fi
  \@tempcnta\z@\@tempcntb\m@ne\def\@citea{}\@cite{\@for\@citeb:=#2\do
    {\@ifundefined
       {b@\@citeb}{\@citeo\@tempcntb\m@ne\@citea\def\@citea{,}{\bf ?}\@warning
       {Citation `\@citeb' on page \thepage \space undefined}}%
    {\setbox\z@\hbox{\global\@tempcntc0\csname b@\@citeb\endcsname\relax}%
     \ifnum\@tempcntc=\z@ \@citeo\@tempcntb\m@ne
       \@citea\def\@citea{,}\hbox{\csname b@\@citeb\endcsname}%
     \else
      \advance\@tempcntb\@ne
      \ifnum\@tempcntb=\@tempcntc
      \else\advance\@tempcntb\m@ne\@citeo
      \@tempcnta\@tempcntc\@tempcntb\@tempcntc\fi\fi}}\@citeo}{#1}}
\def\@citeo{\ifnum\@tempcnta>\@tempcntb\else\@citea\def\@citea{,}%
  \ifnum\@tempcnta=\@tempcntb\the\@tempcnta\else
   {\advance\@tempcnta\@ne\ifnum\@tempcnta=\@tempcntb \else \def\@citea{--}\fi
    \advance\@tempcnta\m@ne\the\@tempcnta\@citea\the\@tempcntb}\fi\fi}
\catcode`@=12

\newcommand{\cD}{{\cal D}}
\newcommand{\cL}{{\cal L}}

\newcommand{\tr}{{\rm Tr}}
\newcommand{\prd}{{\em Phys.\ Rev.\ }  {\bf D}}

\newcommand{\prl}{{\em Phys.\ Rev.\ Lett.\ }}
\newcommand{\np}{{\em Nucl.\ Phys.\ }{\bf B}}
\newcommand{\pl}{{\em Phys.\ Lett.\ }{\bf B}}

\newcommand{\rmp}{{\em Rev.\ Mod.\ Phys.\ }}
\newcommand{\del}{\partial}
\begin{document}
\setcounter{footnote}{1}
\begin{flushright}
MADPH-97-996\\
\end{flushright}
\vspace{14mm}
\begin{center}
\Large{{\bf Effects of the Electroweak Symmetry Breaking Sector in
Rare $B$ and $K$
Decays}}
\end{center}
\vspace{5mm}
\begin{center}
{
Gustavo Burdman}\footnote{e-mail address: burdman@pheno.physics.wisc.edu}\\
*[3.5mm]
{\normalsize\it Department of Physics, University of Wisconsin,}\\ 
{\normalsize
\it Madison, WI 53706, USA}
\end{center}
\vspace{0.50cm}
\thispagestyle{empty}
\begin{abstract}
The effects of the electroweak symmetry breaking sector in rare $B$ 
and $K$ 
decays are considered
in the presence  of new strong dynamics at a large energy scale. 
In the context of the low energy effective lagrangian of the 
symmetry breaking sector,  
we focus on the 
contributions that are not constrained
by bounds on oblique corrections or anomalous triple gauge boson
vertices. 
We find that these have very little 
effect in $b\to s\gamma$, but potentially
large contributions appear in $b\to s\ell^+\ell^-$ and $b\to s\nu\bar\nu$
as well as in $b\to s g$ processes. Deviations
from the standard model in rare $K$ decays such as $K^+\to\pi^+\nu\bar\nu$
and $K_L\to\pi^0\nu\bar\nu$ 
are shown to be correlated 
with the ones in the $B$ modes.  
The distinct resulting pattern of deviations from the standard model in 
rare decays is a characteristic feature of a strongly interacting 
 electroweak symmetry breaking 
sector.

\end{abstract}
\newpage

     \renewcommand{\thefootnote}{\arabic{footnote}}
\setcounter{footnote}{0}
\setcounter{page}{1}

\section{Introduction} 
\vspace{-0.2cm}
If one considers the standard model (SM) as an effective theory, its
remarkable success when confronted with experiment
suggests the possibility that
the dynamics underlying the mechanism for electroweak symmetry breaking
resides at the TeV scale
or above. In particular, in a scenario without a light Higgs boson
the corresponding strong dynamics might manifest itself only at 
very high energies as the ones to be probed by the LHC. 
In some cases, however, the new dynamics can produce sizeable quantum
effects in low energy observables. 
In this letter we study the effects
of a strongly coupled symmetry breaking sector in rare $B$ 
and kaon decays. In
the context of effective field theories, we concentrate on
non-decoupling effects induced by operators not contributing to 
oblique corrections or to shifts in the couplings of three
gauge bosons.
One of the remaining operators, 
which induces $m_t^2$ corrections in charged 
interactions,  
was recently considered in \cite{pich}  where its effects in $R_b$
and  $B^0_d-\bar{B}^0_d$ mixing were estimated. The
logarithmic dependence on the high energy scale $\Lambda$ is used to 
obtain approximate bounds on the corresponding coefficient in the 
low energy expansion. 
We will show that the bounds from these observables still allow for the
possibility of large effects in transitions involving 
flavor changing neutral currents (FCNC), most notably rare $B$ and $K$
decays. 

In the absence of a light Higgs boson the symmetry breaking sector 
is represented by a non-renormalizable effective lagrangian
corresponding to the non-linear realization of the sigma model. 
The essential feature is the spontaneous breaking of the global symmetry
$SU(2)_L\times SU(2)_R\to SU(2)_V$. To leading order the interactions
involving the Goldstone bosons associated with this mechanism and 
the gauge fields are described by
\begin{equation}
\cL_{LO}=-\frac{1}{4}B_{\mu\nu}B^{\mu\nu}
-\frac{1}{2}\tr\left[W_{\mu\nu}W^{\mu\nu}\right]
+\frac{v^2}{4}\tr\left[D_\mu U^\dagger D^\mu U\right] ,
\label{lo}
\end{equation}
where $B_{\mu\nu}$ and  
$W_{\mu\nu}=\del_\mu W_\nu-\del_\nu W_\mu + ig\left[W_\mu,W_\nu\right]$
are the  the $U(1)_Y$ and $SU(2)_L$ field strengths respectively, 
the electroweak scale is $v\simeq 246$~GeV and 
the Goldstone bosons enter through the matrices
$U(x)=e^{i\pi(x)^a\tau_a/v}$. The covariant derivative acting on $U(x)$
is given by $D_\mu U(x)=\del_\mu U(x)+ igW_\mu (x) U(x) - 
\frac{i}{2}g'B_\mu (x) U(x)\tau_3$. To this order there are no free
parameters once the gauge bosons masses are fixed. The 
dependence on the dynamics underlying the strong symmetry breaking
sector appears at next to leading order. A complete set of 
operators at next to leading order includes one operator of 
dimension two and nineteen operators of dimension four 
\cite{longhitano,appelquist}. 
The effective lagrangian to next to leading order 
in the basis 
of Ref.~\cite{longhitano} is given by 
\begin{equation}
\cL_{\rm eff.}=\cL_{LO} + 
\cL'_1 +\sum_{i=1}^{19}\alpha_i\cL_i ~~,
\label{lnlo}
\end{equation}
where $\cL'_1$ is a
dimension two custodial symmetry violating term absent in the heavy
Higgs limit of the SM. If we restrict ourselves to CP invariant
structures, there remain fifteen operators of dimension four. 
The coefficients of some of these operators are constrained by low
energy observables. For instance precision electroweak observables
constrain the coefficient of $\cL'_1$, which gives a contribution
to $\Delta\rho$. The combinations $(\alpha_1+\alpha_8)$ and $(\alpha_1
+\alpha_{13})$ contribute to $S$ and $U$.
The coefficients $\alpha_2$, $\alpha_3$, $\alpha_9$ and $\alpha_{14}$
modify the triple gauge boson couplings and are probed at LEPII at the
few percent level \cite{dieter}.
The remaining operators either contribute to 
oblique corrections to one loop only or do not contribute. 
To the last group belong $\cL_{11}$ and $\cL_{12}$ given that 
their contributions to the gauge boson two-point functions only affect
the longitudinal piece of the propagators. 
Of particular interest is 
the operator $\cL_{11}$ defined by \cite{longhitano}
\begin{equation}
\cL_{11}=\tr\left[\left(\cD_\mu V^\mu\right)^2\right], 
\label{defo11}
\end{equation}
with $V_\mu=(D_\mu U)U^\dagger$ and the covariant derivative acting on
$V_\mu$ defined by $\cD_\mu V_\nu=\del_\mu V_\nu +
ig\left[W_\mu,V_\nu\right]$. The equations of motion for the
$W_{\mu\nu}$ field strength imply \cite{feruglio}
\begin{equation}
\cD_\mu V^\mu = \frac{2i}{v^2}~\cD_\mu J_{w}^{\mu}~, 
\label{equmot}
\end{equation}
where the $SU(2)_L$ current is $J_{w}^\mu=\sum_{\psi}\left(\bar{\psi}_L
\gamma^\mu \frac{\tau^a}{2}\psi_L\right)\tau^a$. The dominant effect 
appears in the quark sector due to the presence of terms proportional to 
$m_t$. After the quark fields are rotated to the mass eigenstate basis, the
operator $\cL_{11}$ can be written as \cite{pich}
\begin{equation}
\cL_{11}=\frac{m_t^2}{v^4}\left\{\left(\bar{t}\gamma_5 t\right)^2
-8\sum_{i,j}V_{ti}^*V_{tj}(\bar{q_i}_L t_R)(\bar{t}_R q_{j L})\right\}
+ \dots 
\label{ffo11}
\end{equation}
where $i,j=d,s,b$, the $V_{ti}$ are Cabibbo-Kobayashi-Maskawa
(CKM) matrix elements and 
the dots stand for terms suppressed by small fermion masses.
Simple inspection of this form of $\cL_{11}$ as a four-fermion operator makes
clear where these corrections will appear. 
In Ref.~\cite{pich} the effects of $\cL_{11}$ in $R_b$ and 
$B^0_d-\bar{B}^0_d$
mixing were studied and found to be binding on the value
of $\alpha_{11}$. There, it was concluded that the mixing limit is slightly
more stringent then the one coming from $R_b$, even when the former is 
plagued
with theoretical uncertainties.
More generally, transitions involving FCNC are likely to be more sensitive
to effects of this kind. In what follows we show that $\cL_{11}$ gives a 
very distinct pattern of potentially large deviations from the SM in 
rare $B$ and $K$ decays. 

Operators of this type are also present 
in scenarios with a light Higgs boson. In this case the SM symmetry
breaking sector is linearly realized. Using 
naive dimensional analysis \cite{nda} we conclude that 
the analogous operators generating
non-oblique corrections and expressible as four-fermion interactions are
of dimension eight or higher and therefore are suppressed by 
additional powers of $v^2/\Lambda^2$. 
Thus the effects we discuss in this
paper are of no significance in the presence of a light Higgs boson. 

In the next section we compute the effects induced in rare $B$ and $K$
processes by the four-fermion operators contained in $\cL_{11}$.  
Additional contributions to these decays from the next-to-leading order
effective lagrangian in (\ref{lnlo})
could come in the form  of anomalous triple gauge boson couplings. We do not
consider these corrections here. 
Moreover, we assume the coupling of the new physics to fermions to be small
and therefore neglect contributions from induced anomalous couplings of 
fermions to gauge bosons \cite{ctofer}, which in any case 
are constrained by the bounds on oblique corrections.  
We summarize and discuss the results in the last section.

\section{Effects in Rare $B$ and $K$ Decays}  
The four-fermion operator in (\ref{ffo11})
\begin{equation}
\cL_{11}=-\frac{8m_t^2}{v^4}\left\{V_{ts}^*V_{tb}\bar{s}_Lt_R\bar{t}_Rb_L
+V_{td}^*V_{tb}\bar{d}_Lt_R\bar{t}_Rb_L+V_{td}^*V_{ts}\bar{d}_Lt_R
\bar{t}_Rs_L
\right\} + \dots
\label{rareo11}
\end{equation}
induces contributions to several FCNC processes when inserted in a top loop,
as shown in Fig.~1.
Among them are 
$b\to q\gamma$, $b\to qZ$ and $b\to q g$ transitions ($q=d,s$) 
as well as the analogous vertices relevant for rare K decays.  
\begin{figure}
\center
\psfig{figure=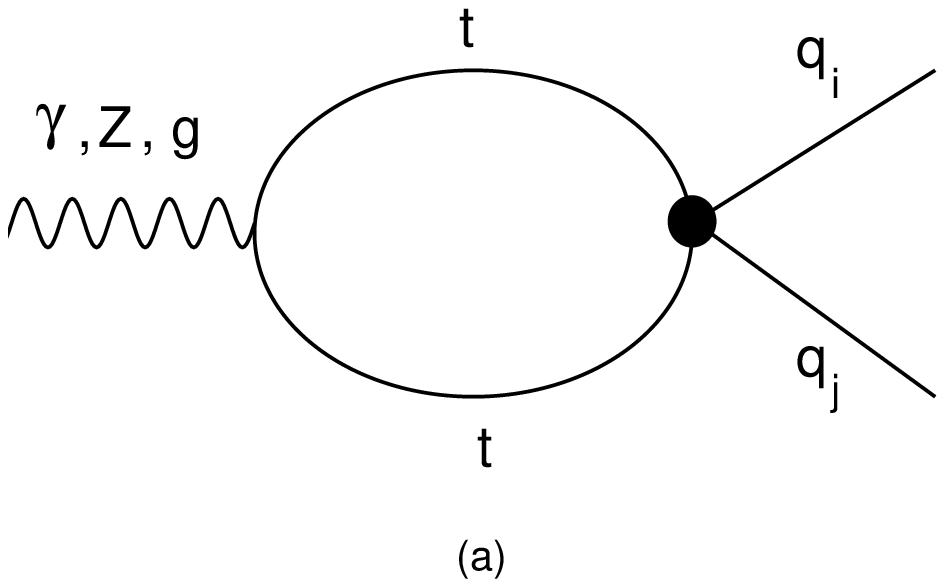,height=1.5in}
\vspace{-1.5in}\hspace{9cm}
\psfig{figure=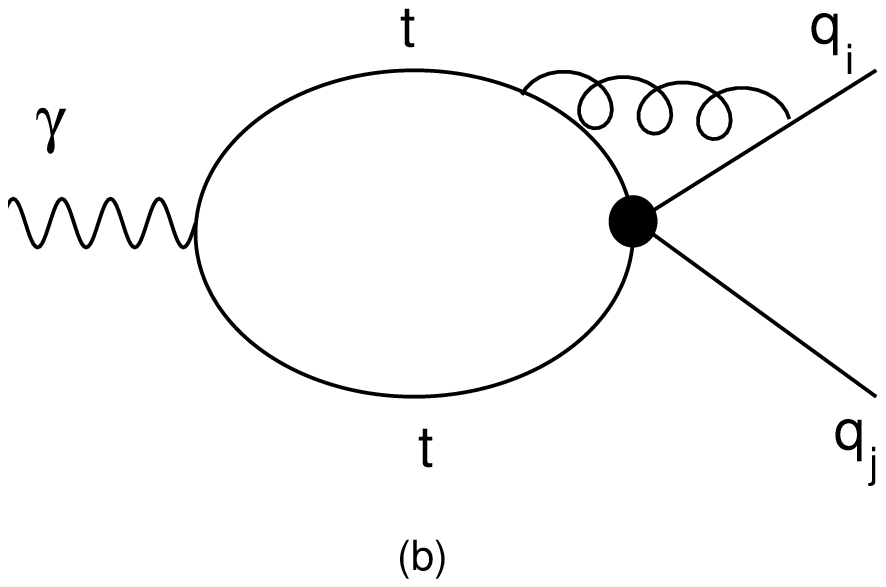,height=1.5in}
\caption{Loop diagrams giving contributions to FCNC processes. The 
black circle denotes the insertion of $\cL_{11}$.
(a) Leading contribution for off-shell photons or gluons, and $Z$ bosons.
(b) One of the two-loop diagrams giving the 
leading contribution to the mixing with the on-shell photon operator
in (\ref{bsgamma}).}
\end{figure}
The insertion of this  non-renormalizable interaction results in a 
logarithmically divergent amplitude when computed using dimensional
regularization. 
The divergences are absorbed by suitable counterterms.  
We will estimate the size of the effects by computing the logarithmic 
dependence
on the high energy scale $\Lambda$. In principle, there could be finite 
effects
coming from the counterterms. However, unless large cancellations occur,
the logarithmic dependence should give a good approximation 
of the effects \cite{nda,burgess}. 
We will come back to discuss this point later in the paper. 
    
We begin with the modification of the $b\to s\gamma$ vertex. 
This can be generically written as
\begin{equation}
\delta\Gamma_{\mu}^{b\to s\gamma}=i\frac{e}{16\pi^2}~V_{ts}^*V_{tb}\left\{
A~\bar{s}_L
\left(\not\!q q_\mu-q^2\gamma_\mu\right)b_L 
+B~m_b\bar{s}_L\sigma_{\mu\nu}b_Rq^\nu\right\} .
\label{bsgamma}
\end{equation}
The second term in (\ref{bsgamma}) gives the on-shell amplitude whereas
the term proportional to $A$ only contributes to the vertex when the photon
is off-shell. 
The computation of the corresponding diagram in Fig.~1~(a) gives 
\begin{eqnarray}
A&=&\frac{8Q_t}{3}\frac{m^2_t}{v^4}\alpha_{11}\log{\left(\frac
{\Lambda^2}{m_t^2}\right)}
\label{offshell} \\
B&=&0~~,
\label{bis0}
\end{eqnarray}
where $Q_t$ denotes the top quark electric charge. 
As it was mentioned before, the value of $A$ in (\ref{offshell}) corresponds
to the logarithmic $\Lambda$ dependence and does not include possible 
finite counterterm contributions. However it gives the correct dependence 
on the
high energy scale $\Lambda$ \cite{burgess}.  
The leading corrections to $b\to s\ell^+\ell^-$ processes 
are therefore governed by $A$. 
From (\ref{bis0}) we conclude that the operator $\cL_{11}$ does not induce
an effect in $b\to s\gamma$ processes at the one loop level. However, next
to leading order corrections induced by the strong interactions will give 
a non-zero value of $B$. The need to go to two loops is characteristic of the 
mixing of four-quark operators with the electromagnetic dipole operator
in (\ref{bsgamma}). A typical contribution of this type is shown in 
Fig.~1~(b).
We compute this mixing after Fierz rearranging  
(\ref{rareo11}) into a product of currents and evolving the corresponding 
coefficient from the scale $\Lambda$ down to $M_W$, where the matching 
to the effective weak hamiltonian is made. 
The dependence with the high energy scale now enters through the factor 
$\frac{\alpha_s}{4\pi}\log{\Lambda^2/M_W^2}$, to be compared with 
(\ref{offshell}) where there is no $\alpha_s$ suppression. 
The effect of the resummation of
these logarithms through the renormalization group equations is small.
The renormalization group running 
from $M_W$ down to $\mu\simeq m_b$ produces the mixing of the four-quark
operator with the second term in (\ref{bsgamma}). This gives, approximately
\begin{equation}
B\approx  -\frac{1}{3}\left[ \sum_{i=1}^{8}h_i\eta^{a_i} \right] \alpha_{11}
\left(\frac{m_t}{v}\right)^2 ,
\label{bsmall}
\end{equation}
where we have neglected a $\simeq10\%$ effect from the running between
$\Lambda$ and $M_W$, we define $\eta\equiv \alpha_s(M_W)/\alpha_s(m_b)$ 
and the 
coefficients $h_i$ and $a_i$ can be found in \cite{nlo}. 
As we will discuss below, this two-loop contribution 
results in a small effect in $b\to s\gamma$ processes. 

Next we consider the $b\to sZ$ vertex. 
The operator (\ref{rareo11}) adds the contribution
\begin{equation}
\delta\Gamma_{\mu}^{b\to sZ}=(-i)\frac{g_W}{c\theta_W}~
V_{ts}^*V_{tb}~\frac{1}{4\pi^2}\frac{m_t^4}{v^4}\alpha_{11}\log{\left(\frac
{\Lambda^2}{m_t^2}\right)}
~\bar{s}_L\gamma_\mu b_L ~, 
\label{bsz}
\end{equation}
with $g_W$ the $SU(2)_L$ gauge coupling and $c\theta_W$ the cosine of the 
Weinberg angle.
The additional factor of $m_t^2$ in (\ref{bsz}) when compared to $q^2$ in 
(\ref{bsgamma}) and (\ref{offshell}) comes from the axial-vector
coupling of the $Z$ boson, which here gives the leading contribution. 

At this point and before estimating the effects of $\cL_{11}$ in rare 
decays, 
we incorporate the constraints on the quantity 
$y\equiv\alpha_{11}\log{\frac{\Lambda^2}{m_t^2}}$, 
characterizing the new physics contributions. 
Implicit in this procedure is the 
assumption that the contributions of finite counterterms 
do not change significantly the size of the contribution.  
Furthermore, there are extensions of the SM where there are no finite
counterterms at the scale $\Lambda$. 
Neglecting
the counterterm contributions allows one to correlate the effects of the 
new dynamics
in various observables, given that these are proportional to $y$.   
The  upper limit for $y$ that is obtained from the 
analysis of $K^0-\bar{K^0}$ and $B^0_d-\bar{B}^0_d$ mixing in \cite{pich}
corresponds to $y<0.4$. A more conservative estimate using a higher 
value for $B_K=0.80\pm 0.20$ \cite{bk} and assuming a $50\%$ uncertainty in 
the value of $|V_{ub}/V_{cb}|$ yields $y<0.50$. A lower limit on $y$ can 
be derived
by comparing the experimental value \cite{lep} 
$R_b=0.2178\pm0.0011$ with the SM
expectation of $R_b^{\rm SM}=0.2158$. Noting that \cite{pich}
\begin{equation}
\frac{\delta\Gamma_b}{\Gamma_b^{\rm SM}}= 4.58\times \frac{1}{4\pi^2}\left(
\frac{m_t}{v}\right)^4 y ~~, 
\label{drb}
\end{equation}
the $95\%$ C.L. lower limit is $y>-0.40$. This limit is largely driven by the 
fact that $R_b$ is larger that the SM prediction by about 
$2\sigma$. 
For instance, had the central value of $R_b$ agreed with the SM, 
the $95\%$ C.L. lower limit would be $y>-0.80$.

The induced vertices of eqns.~(\ref{bsgamma}) and (\ref{bsz}) translate
into deviations from the SM predictions for $B$ decays
governed by the $b\to s\gamma$ and $b\to s\ell^+\ell^-$  transitions. 
They can be 
expressed as shifts in the Wilson coefficients in the weak effective 
hamiltonian 
\begin{equation}
H_{\rm eff.}=-\frac{4G_F}{\sqrt{2}}\sum_{i=1}^{10} C_i(\mu)O_i(\mu),
\label{heff}
\end{equation}
with the operator basis defined in Ref.~\cite{heff}. From eqn.~(\ref{bsmall})
we see that there is a small  shift 
in the coefficient $C_7(M_W)$ of the electromagnetic dipole operator,
$O_7=\frac{e}{16\pi^2}m_b\bar{s}_L\sigma_{\mu\nu}b_R~F^{\mu\nu}$. 
The renormalization group induced mixing of $\cL_{11}$ with $O_7$ 
translates into a shift for this coefficient of the order of 
$\approx -0.05\alpha_{11}$. For instance, for a typical value of $\Lambda=2~$
TeV the bounds on $y$ discussed above give $|\alpha_{11}|<0.1$ which 
gives a  shift of  at most $2\%$ in the 
coefficient $C_7(m_b)$ determining the $b\to s \gamma$
amplitude. 

The coefficient $C_9(M_W)$ of the operator 
$O_9=\frac{e^2}{16\pi^2}(\bar{s}_L\gamma_\mu
b_L)(\bar{\ell}\gamma^\mu\ell)$
is affected almost exclusively by the
shift in the off-shell photon vertex (\ref{offshell}) given that the
contribution to it from the $Z$ exchange is proportional to 
$-1/4+s^2\theta_W\simeq 0$. On the other hand, 
the induced vertex in (\ref{bsz}) produces
a significant shift in $C_{10}(M_W)$, the Wilson coefficient corresponding 
to the operator
$O_{10}=\frac{e^2}{16\pi^2}(\bar{s}_L\gamma_\mu b_L)(
\bar{\ell}\gamma^\mu\gamma_5\ell)$.
We make use of next-to-leading order QCD results for $C_9(m_b)$ 
\cite{nlo}. On the other hand, the coefficient 
$C_{10}(m_b)=C_{10}(M_W)$ and can be found in \cite{heff}.  
The entire analysis above is also valid for $b\to d\ell^+\ell^-$
with the replacements $s\to d$ and $V_{ts}\to V_{td}$, given 
that we have neglected $(m_s/m_b)$ effects.  
In Fig.~2 (solid line) we estimate the size of the effect in 
$b\to (s,d)\ell^+\ell^-$ decays by plotting the ratio
\begin{equation}
R_\ell=\frac{Br(B\to X_{(s,d)}\ell^+\ell^-)}
{~~\;\;Br(B\to X_{(s,d)}\ell^+\ell^-)_{\rm SM}} 
\label{defrat}
\end{equation}
as a function of the quantity $y$.   
We consider a range of values of $y$ compatible with the limits from 
$R_b$ and $B_d^0-\bar{B^0_d}$ mixing
as discussed above.    
We observe that for positive 
(negative) values of $\alpha_{11}$ the branching ratio can be enhanced 
(suppressed) by up to a factor of about two. For instance,
the SM prediction for the  $B\to X_s \mu^+\mu^-$ branching fraction
is about $(5-6)\times 10^{-6}$. 
Thus, the maximum enhancement in $R_\ell$ would
take this  mode to $\approx 1\times 10^{-5}$. However, it should
be noted that the effect 
does not necessarily
affect all {\em exclusive} branching ratios equally. For instance, 
the $B\to K\ell^+\ell^-$ branching fraction
follows $R_\ell$, whereas this is not the case for 
$B\to K^*\ell^+\ell^-$, where
the interplay of the various combinations of 
Wilson coefficients with the helicity 
amplitudes gives a 
quantitatively different answer (e.g. for $y=0.50$ the enhancement 
is $\approx 50\%$, for
$y=-0.50$ the suppression is small).
On the other hand, in these modes the forward-backward asymmetry
for leptons is very sensitive to these type of changes in the 
Wilson coefficients \cite{afb}.   
Current experimental upper limits 
on exclusive modes are already at the $1\times 10^{-5}$ 
level \cite{exp} and sensitivity to SM branching ratios 
will be achieved in the near future. 

\begin{figure}
\center
\hspace{0.20cm}
\psfig{figure=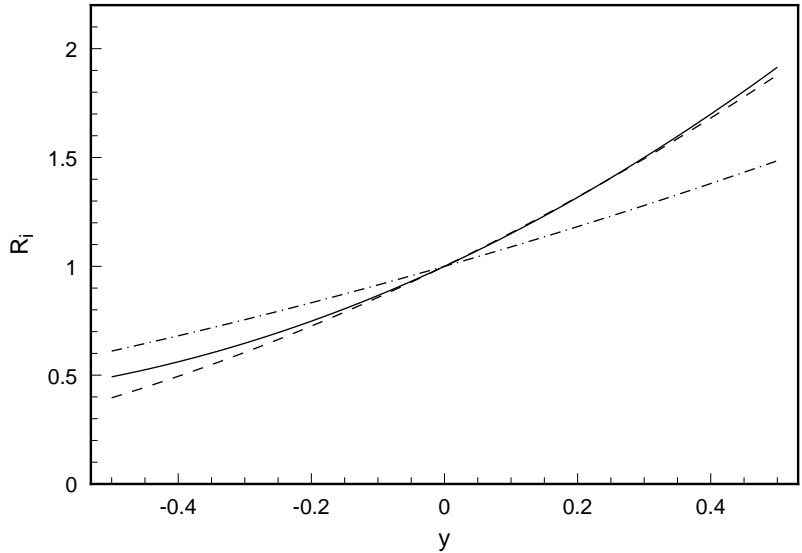,height=3.0in}
\caption{Ratio of the modified branching ratio to the standard model
expectation as a function of $y=\alpha_{11}\log{\frac{\Lambda^2}{m_t^2}}$.
The solid line corresponds to the ratio $R_\ell$ for 
$B\to X_{(s,d)}\ell^+\ell^-$ inclusive decays, the dashed line
to $R_\nu$ for $B\to X_{(s,d)}\nu\bar\nu$ and the dot-dashed line to
$R_g$ for $b\to s\bar s s$ decays.} 
\end{figure}

The modification of the $b\to sZ$ vertex (\ref{bsz}) also induces an effect
in $b\to s\nu\bar\nu$ processes. 
Defining the ratio $R_\nu$  analogously to 
(\ref{defrat})
and plotting it versus $y$ in Fig.~2 (dashed line) 
we see that the effect of $\cL_{11}$ in these decays
approximately follows $R_\ell$. 
The SM expectation is $Br(B\to X_s\nu\bar\nu)
=(4.50\times 10^{-5})$. 
The current $90\%$ C.L. upper limit from LEP
\cite{lepnu} is $7.7\times 10^{-4}$, 
which is still not constraining when compared with $R_\nu$. 

The operator $\cL_{11}$ also induces corrections to the Wilson 
coefficients of QCD penguin operators in the effective weak hamiltonian. 
For instance the modification of the $b\to s g$ vertex due to 
(\ref{rareo11}) is 
\begin{equation}
\delta\Gamma_{\mu}^{b\to s g}=(-i)g\left(\frac{m_t^2}{v^4}\right)
\frac{y}{3\pi^2}
\bar{s}_L\left(\not\!q q_\mu-q^2\gamma_\mu\right)b_L
\label{bsg}
\end{equation}
leading to shifts in the coefficients of the QCD penguin operators 
$C_3$, $C_4$, $C_5$ and $C_6$ defined in Ref.~\cite{heff}, but 
not affecting the gluonic dipole
operator.
In order to estimate the size of the effect we compute the branching fraction  
for the pure QCD penguin process $b\to s\bar s s$. The ratio $R_g$, defined 
analogously to (\ref{defrat}) is plotted in Fig.~2 (dot-dashed line). 
The deviation
with respect to the SM is slightly reduced in this case when 
compared 
to the semileptonic cases. Furthermore, the theoretical
uncertainties associated with the observable exclusive modes are 
rather large
and might obscure any new physics effects \cite{desh}. In any case, 
we see the same correlation with the sign of $y$ as in $R_\ell$ and 
$R_\nu$.  
 
Finally, we turn to study the effects of the operator $\cL_{11}$
in the kaon system. These are induced by the last term in (\ref{rareo11}). 
We focus on the semileptonic decays affected by the modification of the 
$s\to d Z$ vertex, which allows us to concentrate on the theoretically
cleaner modes. For the CP violating mode $K_L\to\pi^0\nu\bar\nu$ the 
ratio to the SM branching fraction, $R_{K_L}$, is identical to 
$R_\nu$ and is therefore given by the dashed line in Fig.~2. 
On the other hand, the equivalent ratio for the $K^+\to\pi^+\nu\bar\nu$
decay depends  mildly on the CP violation parameters $\eta$ and $\rho$
as they appear in the Wolfenstein parametrization \cite{wolf} of the 
CKM matrix,  
as well as on $V_{cb}$. The ratio $R_{K^+}$ is shown in Fig.~3 
(solid line) as a function
of $y$. It can be seen that the pattern of deviation from the 
SM is very similar to that of the neutral mode 
as well as to the 
one observed in rare $B$ decays. The current experimental limit 
on the charged mode is $Br(K^+\to\pi^+\nu\bar\nu)<2.40\times 10^{-9}$ 
\cite{bnl1}
whereas the SM predictions are in the vicinity of $1\times 10^{-10}$. 
Thus it is likely that experiments in the near
future will have sensitivity at the SM level. 
Similar effects will be present in other decay modes, for instance 
$K_L\to\mu^+\mu^-$, 
$K^+\to\pi^+\mu^+\mu^-$, etc.However, these decays are  
contaminated by 
potentially large long distance contributions \cite{longd}.

\section{Summary and Discussion}
In a scenario without a light Higgs boson the electroweak sector is 
non-linearly realized and most likely underlied by strong dynamics. 
We have studied the effects of such a scenario in rare $B$ and $K$ decays. 
These originate almost exclusively in one operator in the next-to-leading
order effective lagrangian for the Goldstone bosons, given that its
effects are proportional to $m_t^2$. 
The dependence on the high energy
scale $\Lambda$ resulting from the insertion of the operator $\cL_{11}$ given 
in (\ref{rareo11}) in FCNC loops is given by the factor 
$y=\alpha_{11}\log{\frac{\Lambda^2}{m_t^2}}$. 
We further assume this dependence
to be the leading contribution to the amplitudes. In doing so we neglect
possible finite contributions from counterterms coming from the matching 
at the scale $\Lambda$. This allows us to correlate the effects in  
several observables. We have shown that, even after imposing the bounds
on $y$ from $R_b$ and $B^0_d-\bar{B^0_d}$ mixing, large effects
on rare $B$ and $K$ decays remain, as can be seen in Figs.~2
and~3. A very distinct pattern of deviations from the SM emerges. 
Positive values of $y$ give similar enhancements of up to a
factor of almost two in 
$B\to X_{(s,d)}\ell^+\ell^-$, $B\to X_{(s,d)}\nu\bar\nu$, $b\to s\bar s s$
and other QCD gluonic penguin decays, $K^+\to\pi^+\nu\bar\nu$ and 
$K_L\to\pi^0\nu\bar\nu$, whereas negative values of $\alpha_{11}$ imply
a similar suppression of up to a factor of two in all these processes. 
Other modes, such as $B_{(s,d)}\to\ell^+\ell^-$, 
$K_L\to\ell^+\ell^-$ and $K^+\to\pi^+\ell^+\ell^-$, are similarly affected
but we concentrated on those processes that are both theoretically
clean and most accessible to present and future experiments. 
The effects of the operator $\cL_{11}$ in 
on-shell photon processes as $b\to s\gamma$, $s\to d\gamma$, etc. are 
only induced by the two-loop strong interaction mixing  and are expected
to be rather small when compared with those in the leptonic and semileptonic
modes for the same values of the parameter $y$. 

Both the correlation among different processes as well as the size of
the effects are well approximated by the logarithmic behavior
as long as there are no large cancellations due 
to counterterms. Moreover, 
in extensions of the SM where the new dynamics associated
with the electroweak symmetry breaking does not couple 
appreciably to fermions, 
there will be no counterterms and the 
observed correlation among processes is not
just an expected pattern but a quantitatively valid statement 
stemming from the calculation of the effects. 
This is consistent with our assumption of no anomalous couplings of 
fermions to gauge bosons.

\begin{figure}
\center
\hspace{0.25cm}
\psfig{figure=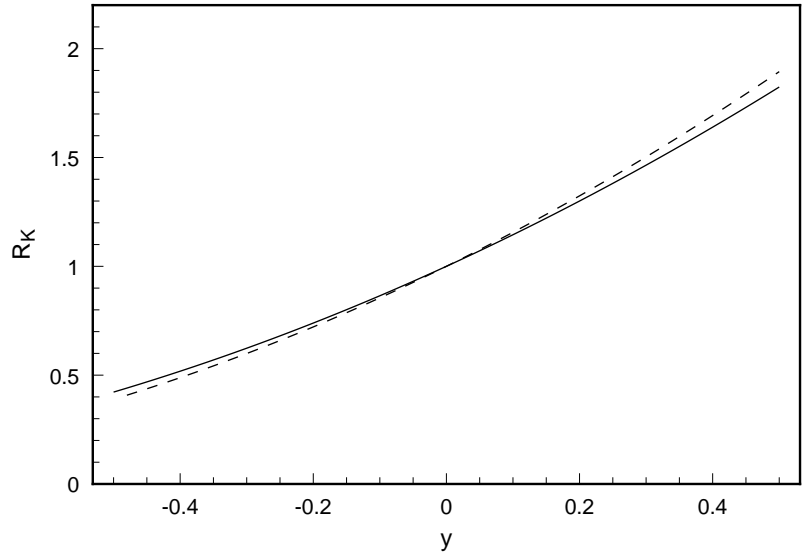,height=3.0in}
\caption{Ratio of the modified branching ratio to the standard model
expectation for $K^+\to\pi^+\nu\bar\nu$ (solid line) and 
$K_L\to\pi^0\nu\bar\nu$ (dashed  line).}
\end{figure}

Regarding the size of the coefficient $\alpha_{11}$ we note that, for 
a high energy scale of e.g. $\Lambda = 2$~TeV, its value is allowed to be 
$|\alpha_{11}|< 0.1$. We have not made an attempt to estimate the size of
this coefficient in extensions of the SM. In the scenario considered in this 
work, the new dynamics is strongly coupled and the coefficients of the 
effective lagrangian can be large.    

Finally, we point out that this is one more example of deviations occurring
at the same level in both rare $B$ and $K$ decays. This is characteristic
of new physics entering in loop-induced FCNC as long as the new dynamics
couples equally to all generations. In this way, $m_t^2$ effects associated
with the electroweak symmetry breaking affect similarly FCNC with 
external down-type quarks. 
Therefore the 
experimental
information from both types of processes is essential in disentangling
the source of the effect in scenarios like the present one. 
Experiments designed to achieve sensitivity to the SM branching fractions, 
such as leptonic and hadronic $B$ factories and high intensity kaon 
experiments can have interesting consequences for the physics
at the high energy scale $\Lambda$ even when this can only be directly
probed at the LHC.

\vskip1.50cm
\noindent
{\bf Acknowledgments}

\noindent
The author thanks German Valencia and Dieter Zeppenfeld for helpful
discussions and  critical readings of the manuscript.
This work was supported  by the U.S.~Department of Energy 
under  
Grant No.~DE-FG02-95ER40896 and the University of 
Wisconsin Research Committee with funds granted by the Wisconsin 
Alumni Research Foundation.

\end{document}